\documentclass[aps, pra, reprint, superscriptaddress, nofootinbib]{revtex4-2}

\usepackage{amsmath, amssymb}
\usepackage{graphicx}
\usepackage[colorlinks=true, linkcolor=blue, citecolor=blue, urlcolor=blue]{hyperref}
\usepackage{braket}
\usepackage{cancel}
\usepackage[caption=false]{subfig}

\begin{document}

\title{Tracking Quantum State Collapse/Decoherence in Real Time via a Superposition Trap}

\author{Hardeep Singh}
\email{hardeep.singh@u.northwestern.edu}
\affiliation{Department of Physics and Astronomy and Center for Fundamental Physics, Northwestern University, Evanston, Illinois, 60208, USA}

\author{Tim Kovachy}
\affiliation{Department of Physics and Astronomy and Center for Fundamental Physics, Northwestern University, Evanston, Illinois, 60208, USA}
% Add additional authors and affiliations as needed

\date{\today}

\begin{abstract}
% A brief summary of your work—what problem you address, what you propose, and its significance.
We propose a novel method for probing quantum state collapse and decoherence in real time using a mechanism we term the \textit{superposition trap}. This approach exploits the continuity equation in quantum mechanics to engineer a configuration that spatially confines only coherent superpositions, while allowing decohered or collapsed states to escape. By monitoring this leakage, the dynamics of collapse processes—whether environmentally induced or intrinsic to objective collapse models—can be experimentally accessed without disturbing the coherent evolution itself. We outline a concrete implementation using an atom interferometer with strontium atoms, where internal-state-selective operations enable physical separation of collapsed components. This technique offers a new experimental avenue to test collapse models, measure decoherence rates, and investigate the quantum-to-classical transition.
\end{abstract}

\maketitle

\section{Preamble}

The measurement problem in quantum mechanics remains one of the most profound unresolved issues in modern physics \cite{von2018mathematical,maudlin2019philosophy,bub1999interpreting}. At its core lies the question of whether wavefunction collapse is a real, dynamical physical process or merely a convenient description arising from entanglement with an uncontrolled environment—a question central to our understanding of quantum reality \cite{zeh1970interpretation,schlosshauer2007decoherence}. Numerous interpretations, from Copenhagen to many-worlds, attempt to explain how and why measurement outcomes appear definite despite underlying superpositions, but experimental access to the collapse process itself has remained elusive \cite{wallace2012emergent, kent2009one}.

Recent decades have seen the rise of objective collapse models such as the Ghirardi-Rimini-Weber (GRW) \cite{ghirardi1986unified}, Continuous Spontaneous Localization (CSL) \cite{ghirardi1990markov}, and Di\'{o}si-Penrose models \cite{diosi1989models, penrose1996gravity}, which posit that the collapse is a stochastic physical process that modifies the unitary evolution of quantum mechanics . While these models make testable predictions, experimental bounds are still far from definitively ruling them out \cite{bassi2013models, bassi2003dynamical, carlesso2022present}. Decoherence theory, on the other hand, provides a framework to explain the suppression of interference via entanglement with an external environment, but it does not solve the problem of definite outcomes \cite{schlosshauer2004decoherence, zurek1981pointer, zurek1982environment, joos2013decoherence}.

Monitoring decoherence and collapse rates in real time can be useful, as it facilitates the empirical identification of factors that influence these processes. Such capability would provide a powerful tool for elucidating the physical mechanisms underlying quantum state reduction. In this work, we propose a novel approach based on what we call the \textit{superposition trap}, which exploits structural features of the Schrödinger equation—particularly the continuity equation—to isolate and monitor quantum states that retain coherence. Our method enables the construction of a spatial or internal-state “trap” that confines only coherent superpositions, allowing decohered or collapsed states to escape and be detected. This provides a new route to experimentally probe the dynamics of collapse processes with minimal theoretical assumptions.

This paper is structured as follows. We begin by reviewing the theoretical underpinnings of quantum continuity and the measurement problem, emphasizing the role of the continuity equation in unitary quantum evolution and its tension with wavefunction collapse. Building upon this foundation, we introduce the concept of the \textit{superposition trap}—a mechanism that selectively confines coherent superpositions based on interference and continuity conditions. We then develop the physical intuition behind the trap and present an idealized optical implementation using an interferometric arrangement. Finally, we propose a practical realization using atom interferometry, detailing how internal-state-selective transitions in strontium atoms can be used to isolate and monitor decohered or collapsed components.

\section{Background Theory}

We will first lay the theoretical ground that lies at the heart of the \textit{superposition trap}, which is the continuity constraint. In fluid dynamics, the continuity equation is a key element constraining the flow of fluids. Likewise in quantum mechanics we will find that the continuity equation gives us a constraint on the evolution of quantum states \cite{landau1987fluid, landau2013quantum}. The fact that the probability density current is non-linear in the wavefunction implies that superposition plays a non-trivial role \cite{madelung1927quantum, bohm1952suggested}.

\subsection{Continuity in Quantum Evolution}\label{sec: continuity}
% Brief discussion of unitary evolution and the role of Schrödinger equation.

The continuity equation imposes a significant constraint on the quantum dynamics. When one looks at the Schrödinger equation, this constraint is not immediately apparent. Nevertheless, most standard quantum textbooks derive the continuity equation for the probability density—a result that reveals an important structural feature of quantum theory. Notably, the notion of probability density continuity in quantum mechanics takes a rather strange form, as the probability density current does not depend explicitly on the probability density \cite{sakurai2020modern, griffiths2018introduction, landau2013quantum, cohen2019quantum}. The continuity equation in canconical form is written as

\begin{equation}\label{eqn:continuity}
    \frac{\partial \rho(\mathbf{r}, t)}{\partial t} + \nabla \cdot \mathbf{j}(\mathbf{r}, t) = 0
\end{equation}
for the the velocity independent potentials, the probability current density $\mathbf{j}(\mathbf{r}, t)$ is given by
\begin{equation}
\mathbf{j}(\mathbf{r}, t) = \frac{\hbar}{2mi} \left[ \psi^*(\mathbf{r}, t) \nabla \psi(\mathbf{r}, t) - \psi(\mathbf{r}, t) \nabla \psi^*(\mathbf{r}, t) \right]
\label{eqn: probabiity current}
\end{equation}
Taking the volume integral of Eq.~(\ref{eqn:continuity}) reveals that the change in the total probability within a volume is entirely determined by the flux of the probability current through its boundary. 
\begin{equation}
    \frac{\partial}{\partial t} \int{\rho(\mathbf{r}, t) \, dV} = - \oint \mathbf{j}(\mathbf{r}, t)\cdot d\mathbf{S}
\end{equation}
As per Eq. \ref{eqn: probabiity current}, $\mathbf{j}(\mathbf{r}, t) = 0$ wherever $|\psi(\mathbf{r}, t)|^2 = 0$. 

This leads to an intriguing implication: a closed surface defined by $|\psi(\mathbf{r}, t)|^2 = 0$ completely traps the probability density within it. This is especially interesting in quantum mechanics because we know that no finite potential barrier can fully confine a quantum particle—the wavefunction can always tunnel out. Perfect confinement typically demands an infinite potential, which is unphysical. Yet, the continuity equation offers a subtler mechanism—one not tied to potential but to the structure of the wavefunction itself—for fully trapping probability density. It is to be noted that the null points (position where probability density is zero) can evolve in time \cite{singh2022study}.

One might wonder whether the continuity equation in the form above applies only under special conditions, potentially limiting its practical relevance. To address this, we turn to the Feynman path integral formulation. A simple rearrangement of the path summation shows that if a boundary exists where the probability density vanishes, then all paths crossing that boundary contribute zero net amplitude. In other words, the paths that traverse a surface where $|\psi|^2 = 0$ are exactly canceled out in the sum over all other paths. This reinforces the earlier conclusion: probability density within such a boundary remains constrained. Such a closed boundary can also be referred to as a *probability density bubble*~\cite{singh2022study}. Appendix \ref{Appendix: continuity path} presents that the probability density bubble constraint should be true to any modification to Quantum Mechanics as long as it is compatible with  the Feynman path integral formalism.

\subsection{Measurement and Collapse}
% Introduce the problem of measurement, Born rule, and standard interpretations.
The measurement problem has remained a central conceptual challenge in quantum mechanics for nearly a century \cite{maudlin2019philosophy}. Despite extensive theoretical developments, no consensus has emerged regarding the precise mechanism by which a quantum system, initially described by a superposition of states, yields a single, definite outcome upon measurement \cite{bub1999interpreting, wallace2012emergent, bassi2013models}. The difficulty stems from the fact that any realistic measurement apparatus necessarily involves coupling the quantum system to an environment with an enormous number of degrees of freedom—often on the order of Avogadro's number\cite{joos2013decoherence, schlosshauer2004decoherence}. This renders it effectively impossible to identify a unique point in time or space where the transition from superposition to definite outcome occurs. The resulting disconnect between the formalism of unitary evolution and the emergence of classical outcomes remains one of the most intriguing and unresolved puzzles in quantum theory.

\begin{figure}[h]
    \centering
    \includegraphics[width=0.6\linewidth]{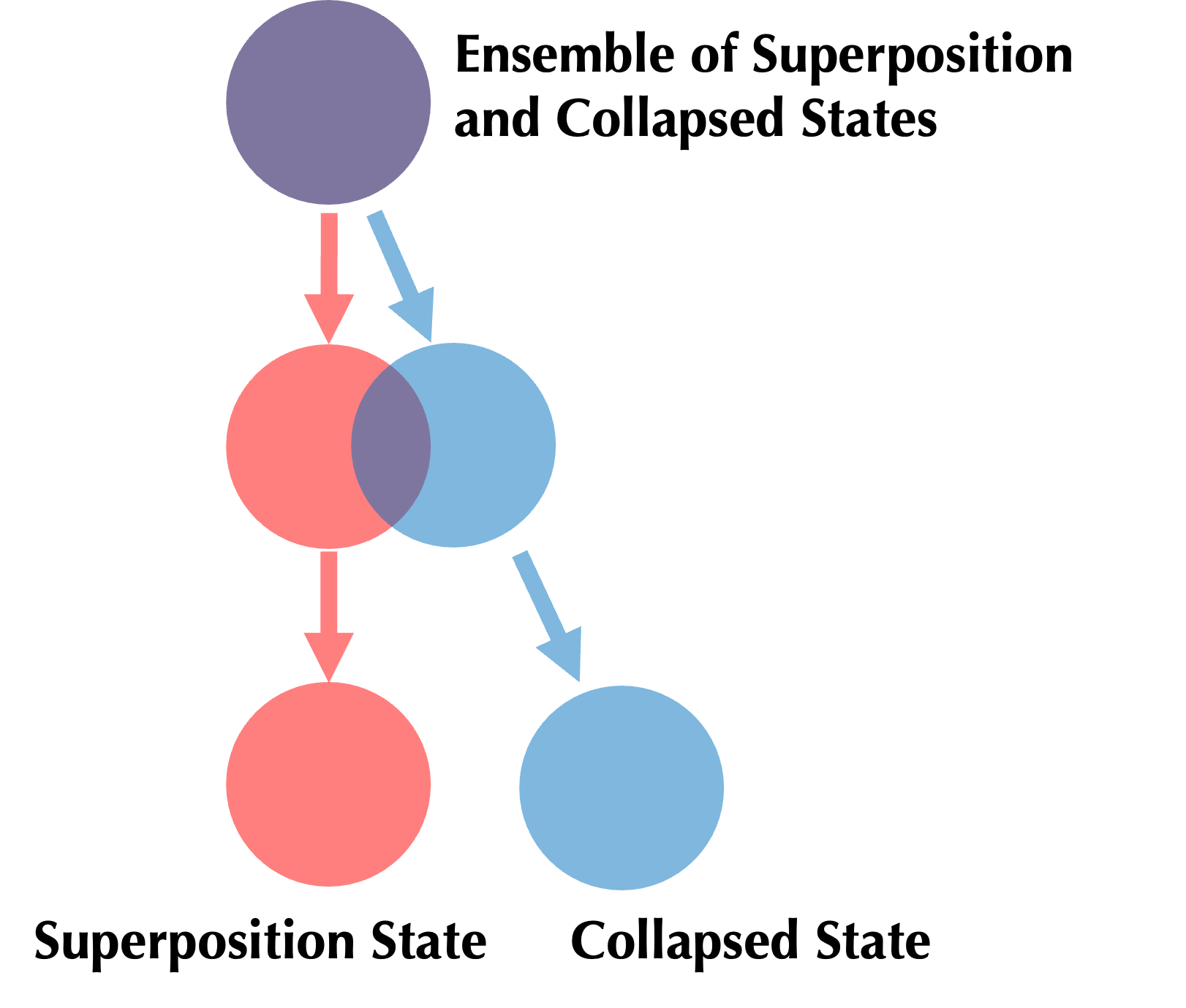}
    \caption{This figure demonstrates the abstract goal of a superposition trap, physically separating the collapsed state from the superposition state.}
    \label{fig: Abstrat separation}
\end{figure}

To advance beyond speculation, it can be useful to develop experimental or theoretical probes that can access the dynamics of collapse or decoherence. This inspires one to devise a strategy that allows for monitoring the process in an ensemble without itself inducing collapse. In other words, the goal is to isolate the superposition states in the ensemble from the post-collapse states and perform observations solely on the latter. Figure \ref{fig: Abstrat separation} illustrates the abstract goal, that is to separate out the superposition state from the collapsed state. This raises the fundamental question: Is it possible to construct such a system within the framework of standard quantum theory?

To address this, we begin by re-examining the foundational postulates of quantum mechanics \cite{sakurai2020modern}. While various formulations differ in presentation, we enumerate here only those postulates directly relevant to our discussion.
\begin{enumerate}
    \item \textbf{State Postulate:} The state of a quantum system is described by a normalized vector $|\psi\rangle$ in a complex Hilbert space $\mathcal{H}$.
    
    \item \textbf{Unitary Evolution Postulate:} The evolution of a closed quantum system is described by a unitary operator $U$ acting on the state:
    \[
    |\psi'\rangle = U |\psi\rangle.
    \]

    \item \textbf{Time Evolution Postulate:} The time evolution of a closed quantum system is governed by the Schrödinger equation\footnote{Almost all textbooks will put the second and third postulate under one bracket. However, here we desire to think of evolution not just in time, but through any general parameter or process. The requirement of unitary ensures that the size of the vector remains fixed in the Hilbert Space. Hence, we listed them as separate postulates}: 
    \[
    i\hbar \frac{d}{dt}|\psi(t)\rangle = \hat{H}|\psi(t)\rangle.
    \]

    \item \textbf{Observable Postulate:} Every observable is represented by a Hermitian operator $\hat{A}$; the possible outcomes of a measurement are its eigenvalues.

    \item \textbf{Measurement Postulate:} The probability of obtaining eigenvalue $a$ in a measurement is $|\langle a | \psi \rangle|^2$. After measurement, the system collapses to the corresponding eigenstate $|a\rangle$.
    
\end{enumerate}
The first and second postulates of quantum mechanics establish that the state of a quantum system is described by a vector in a Hilbert space, and that the norm of this vector is preserved under evolution. The third postulate specifies the time evolution of the system, governed by the Schrödinger equation, and implicitly requires that this evolution be continuous and unitary. The fourth postulate introduces the mathematical machinery for extracting physical information via observables represented by Hermitian operators. Finally, the fifth postulate—the so-called projection or collapse postulate—states that, upon measurement, the quantum state instantaneously collapses to an eigenstate of the measured observable.

Closer scrutiny reveals an apparent inconsistency between the third and fifth postulates, particularly in the context of measurements in the position basis. The fifth postulate allows for a wavefunction initially in a superposition of spatially separated states to collapse into one of the position eigenstates, even when those locations are macroscopically distinct and the wavefunction has vanishing probability amplitude in the intermediate region. Such a discontinuous transition is not permitted under the third postulate, which enforces continuous unitary evolution through the Schrödinger equation (cf. Sec \ref{sec: continuity}). This inconsistency suggests that either the standard time-evolution postulate is incomplete when it comes to describing measurement, or that the collapse process lies outside the domain of unitary evolution altogether \cite{bassi2013models}. Thus, we arrive at an Archimedean point\footnote{According to The Oxford dictionary of philosophy, Archemedian point is defined as: ``Metaphor derived from Archimedes's alleged saying that if he had a fulcrum and a lever long enough, he could move the earth. The Archimedean point is a point ‘outside’ from which a different, perhaps objective or ‘true’ picture of something is obtainable." \cite{blackburn2005oxford}
}: collapse processes must evade the constraints imposed by the continuity and unitarity of standard time evolution. Understanding this evasion is central to making progress on the measurement problem.

\section{Concept of the Superposition Trap}
\subsection{Trap Theory}\label{sec: Trap Theory}
The preceding discussion provides a compelling basis for using continuity as a distinguishing feature between superposition and collapsed or decohered quantum states. Specifically, continuity—enshrined in the unitary time evolution governed by the Schrödinger equation—can serve as a tool to isolate and manipulate only those states that retain quantum coherence. This motivates the concept of a superposition trap: a spatial configuration that selectively confines superposed quantum states while allowing collapsed states to escape.

\begin{figure}[h]
    \centering
    \includegraphics[width=0.8\linewidth]{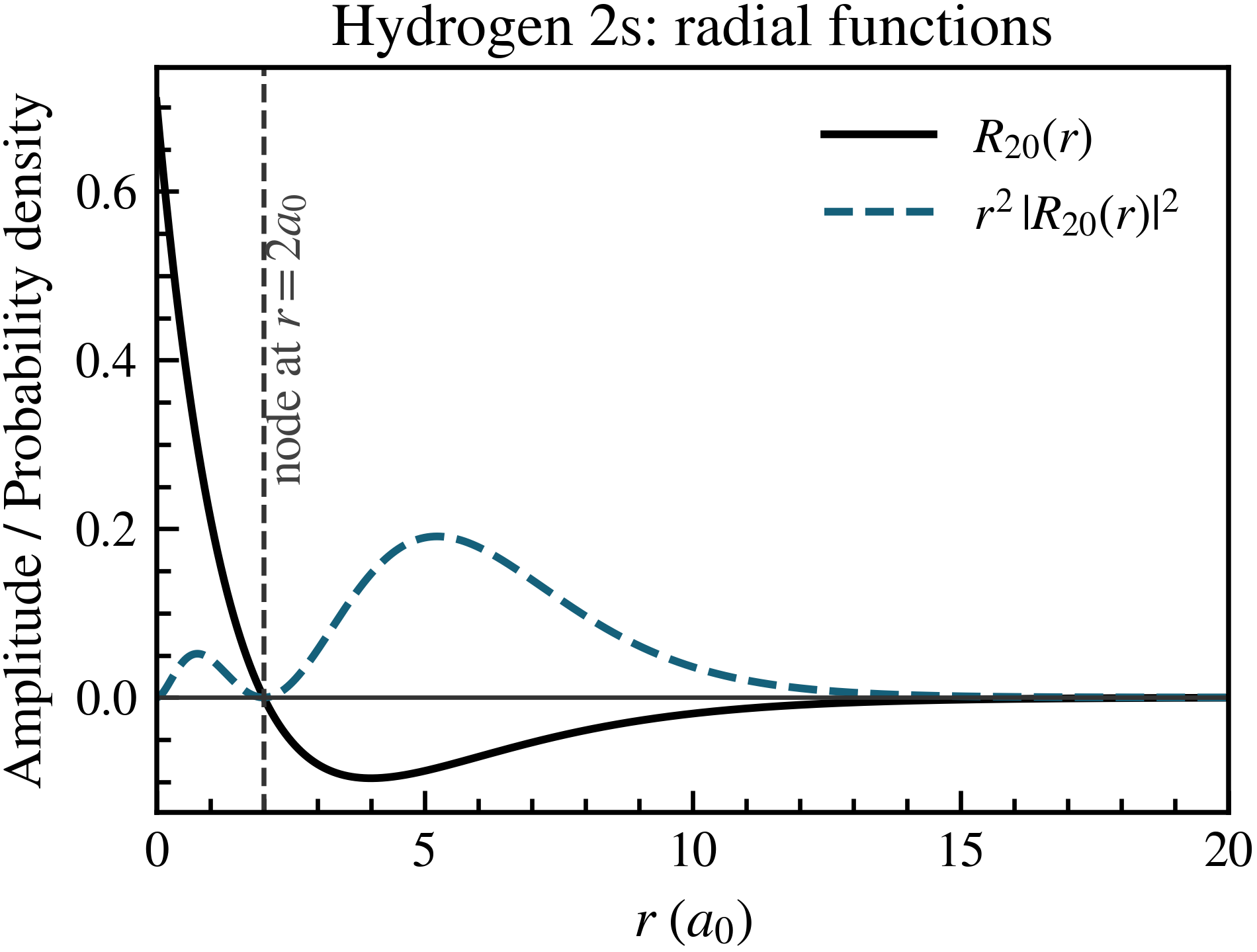}
    \\[1em]
  \includegraphics[width=\linewidth]{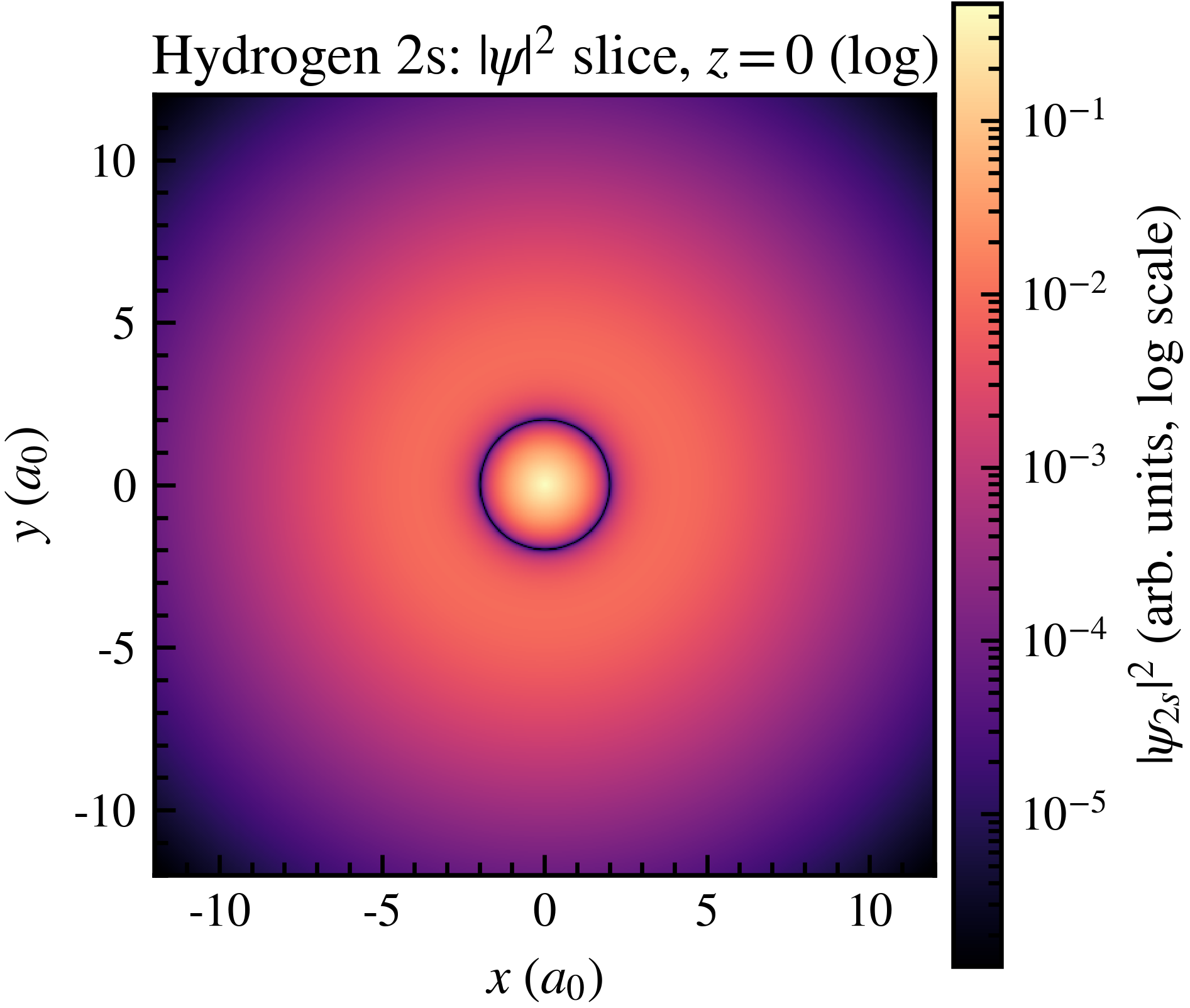}
    \caption{ Top figure shows the radial wavefunction $R_{20}(r)$ and radial probability density
  $r^{2}\lvert R_{20}(r)\rvert^{2}$ of the hydrogen $2s$ state
  (atomic units). The vertical dashed line marks the radial node
  at $r = 2 a_{0}$.
  Bottom figure shows logarithmically scaled density plot of
  $\lvert \psi_{2s}(x,y,z\!=\!0) \rvert^{2}$ for the hydrogen $2s$ orbital
  in the plane $z=0$ (atomic units). Node can be clearly seen as a dark ring.}
    \label{fig:Hydrogen}
\end{figure}

To illustrate this idea, consider a stationary wavefunction characterized by a localized region of non-zero probability density, bounded entirely by surfaces where the wavefunction destructively interferes, resulting in nodes of zero probability density. We refer to such boundaries as null boundaries \cite{singh2022study}. According to the continuity equation and unitary evolution (cf. Sec.~\ref{sec: continuity}), if the null boundary is maintained, the total probability enclosed within must remain conserved in time. Such conditions naturally arise in bound quantum systems; for instance, the $S$-orbitals of the hydrogen atom with principal quantum number $n > 1$ exhibit spherically closed nodal surfaces where the probability density vanishes \cite{pilar2001elementary}. Figure \ref{fig:Hydrogen} plots the $2S$ orbital of the hydrogen atom, for which the closed nodal surface is at $2a_0$.

\begin{figure}[h]
    \centering
    \includegraphics[width=0.6\linewidth]{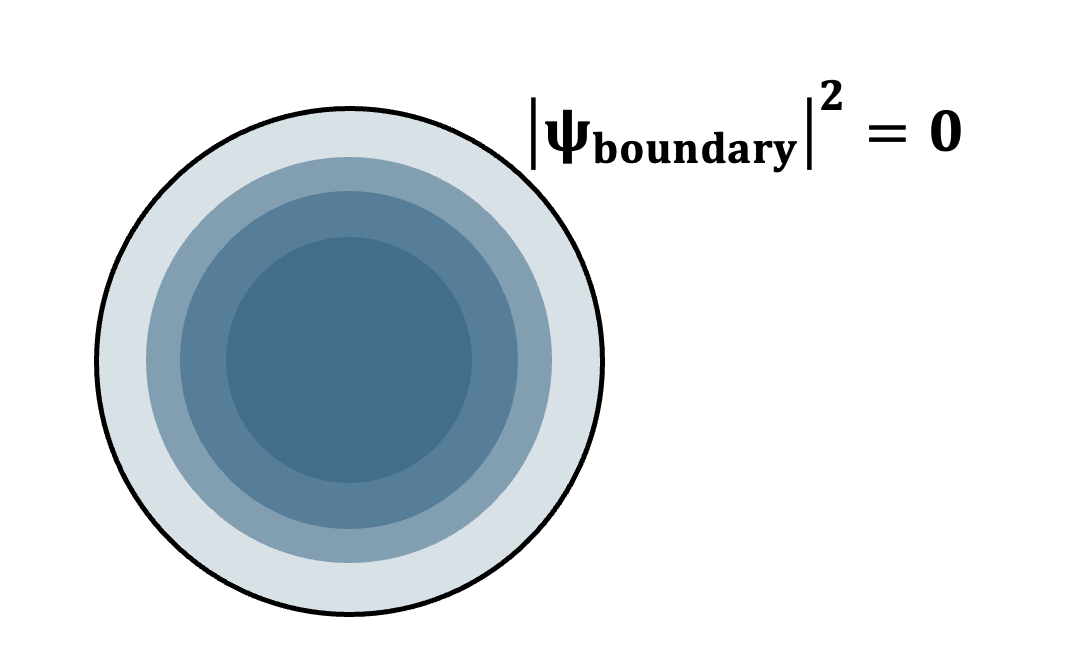}
    \caption{Abstract figure represents the continuity constraint. If the surface is made of zero probability points then the probability density is contained by the closed surface}
    \label{fig: trapped probability}
\end{figure}

Now, suppose we engineer a quantum state such that its entire probability density is localized within a closed volume defined by a surface $S$, and that destructive interference ensures vanishing probability current through all open apertures on $S$. Under these conditions, the continuity equation guarantees confinement of the probability density within $S$. As illustrated in Fig. \ref{fig: trapped probability}, the border of the circle has the probability density zero. This implies that the probability density inside the circle will remain contained as long the boundary condition of the probability density being zero is met.

However, if the quantum state collapses or undergoes decoherence, the coherent superposition necessary for destructive interference at the surface $S$ is lost. Consequently, the wavefunction can acquire non-zero amplitude at previously nodal boundaries, allowing probability density to leak through the openings. In this way, surface $S$ no longer acts as a confining boundary for the decohered or collapsed state.

Thus, the surface $S$ selectively confines quantum states that preserve interference—i.e., coherent superpositions—while permitting incoherent, post-collapse components to escape. This mechanism forms the basis of what we term the \textit{superposition trap}: a structure or configuration that differentiates between coherent and incoherent quantum states based on their continuity properties, allowing for potential detection or manipulation of collapse dynamics. For a single particle, if the particle is observed to have escaped from the trap, then it is established that the particle has collapsed. Thus, no population remains in the superposition trap.  For an ensemble of particles, one can monitor the subset of particles that has collapsed and escaped, while other particles remain coherently trapped. Even for the single particle case, monitoring the time at which it escapes can provide information about when the collapse occurs.

\subsubsection{Optical Trap}
In quantum optics and quantum information science, interferometric setups such as the Mach-Zehnder interferometer often serve as ideal testbeds\cite{wheeler1978past, zeilinger1999experiment, scully1982quantum, grangier1986experimental, mandel1999quantum, Home_Whitaker_1992}. Remarkably, a variation of the superposition trap can be implemented using such an arrangement. The inspiration can be taken from the ``Elitzur–Vaidman bomb tester", where we test the existence of a working bomb in an interferometer arm \cite{elitzur1993quantum, kwiat1995interaction}. The bomb detection uses destructive interference at one of the detectors, essentially silencing it. In the presence of the bomb, the coherence between arms is lost, spoiling the destructive interference. Thus, the previously silenced detector is able to detect a signal.

The bomb is a collapse mechanism. In the original thought experiment, the bomb was always present. However, in general, collapse happens at random. If we model the collapse as the presence of a working bomb, then we can consider a picture where the working bomb randomly appears in an interferometer arm. A natural question is: Can we somehow find out when the bomb appears? Perhaps, if one can turn the ``Elitzur–Vaidman bomb tester" experiment into a repetitive loop.  Then, monitoring when a signal appears at the otherwise silenced detector conveys information about when the bomb materialized.

\begin{figure}[h]
    \centering
    \includegraphics[width=\linewidth]{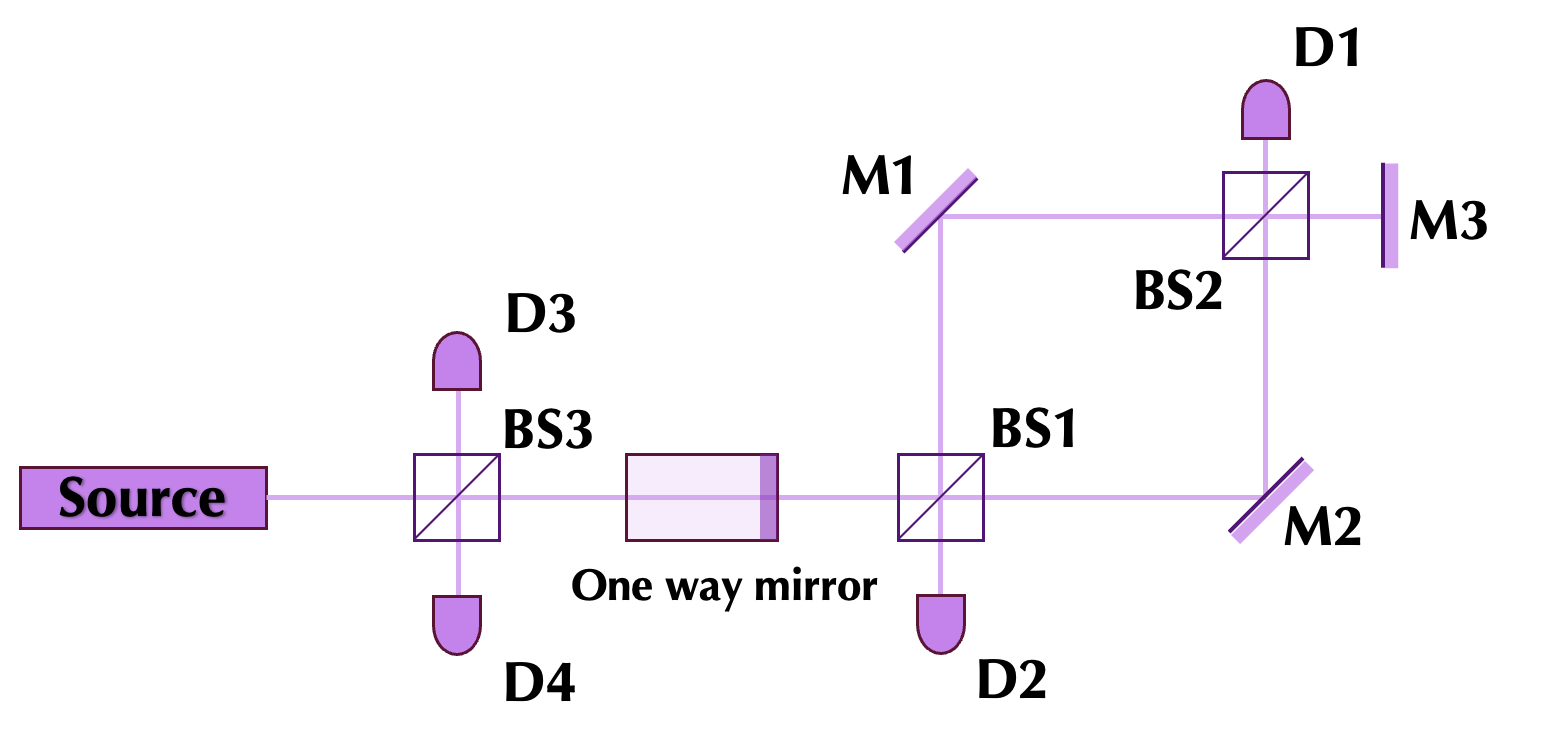}
    \caption{The schematic represents the schematic of the optical trap. In the ideal case the light will be trapped between $M3$ and one way mirror, as long as it remains coherent. On the other hand, light can be detected at $D1$ and $D2$ when there is decoherence while traversing through the interferometer. Thus, $D1$ and $D2$ only detects the collapsed state.}
    \label{fig:Optical Trap}
\end{figure}

Figure~\ref{fig:Optical Trap} illustrates a possible realization. The setup consists of symmetric beam splitters (BS1 and BS2) and mirrors. Polarization-maintaining elements are assumed to ensure balanced interference. In the ideal case, where all optical paths are precisely matched and coherent, photons entering the interferometer from the source interfere destructively at the detector after BS2. Consequently, the probability amplitude at the detector is zero, and all photons end up at the mirror are reflected back into the interferometer. A similar destructive interference occurs at the input port, such that photons circulate within the interferometer and do not return to the source.

To complete the trap, a one-way optical element—a hypothetical optical diode or engineered one-way mirror—is introduced at the input. This element allows photons to enter the interferometer from the source but prevents them from exiting in the reverse direction.  In this configuration, the mirrors and beam splitters collectively define the confining boundary surface $S$ discussed in Sec.~\ref{sec: Trap Theory}.

The key insight is that this configuration traps only coherent superpositions: interference conditions ensure that ideal, uncollapsed wavefunctions remain confined within the interferometer. However, if a photon decoheres or collapses while propagating through the arms—due to environmental interaction or a measurement-like event—it no longer exhibits perfect interference. In that case, the destructive interference at the detector port is lost, allowing a finite probability for the photon to reach the detector. Thus, only collapsed or decohered states are able to escape, while coherent superpositions remain trapped within the mirror.

This optical analog provides a concrete realization of the superposition trap, wherein coherence is used as a control parameter for spatial confinement. Although ideal destructive interference requires precise alignment and stability, this design lays the groundwork for more advanced implementations capable of probing collapse dynamics via spatial leakage of probability.

\subsubsection{Atomic Trap}
The Mach-Zehnder scheme provides a valuable conceptual foundation for realizing a superposition trap. However, its implementation using optics poses practical challenges, chiefly due to the requirement of a one-sided mirror—a component that, to the best of the authors' knowledge, does not exist in standard photonic systems. In contrast, atom interferometers offer a natural alternative, wherein beam splitters and mirrors are replaced by appropriately timed sequences of laser pulses \cite{tino2014atom, cronin2009optics}. Since these pulses are applied temporally rather than spatially, the need for a one-way boundary condition is inherently circumvented.

In this atom-based analog, the goal is to engineer an interferometric sequence in which atomic wave packets interfere constructively in one arm—designated the \textit{main arm}—and destructively in the other, which we term the \textit{collapsed arm}. Unlike in photonic setups, if atoms in the collapsed arm are not actively removed, continued pulse sequences will re-mix the states, negating the distinction. Therefore, a key requirement is the physical separation of atoms in the collapsed arm before further evolution.

This separation can be achieved using internal-state-selective optical transitions. Atoms can be prepared in a specific internal state, and after interferometric evolution, those in the undesired (collapsed) arm can be selectively addressed and removed or displaced. The remaining atoms in the main arm continue coherent evolution, while those removed from the collapsed arm act as a statistical log of decoherence events.

\begin{figure}[h]
    \centering
    \includegraphics[width=\linewidth]{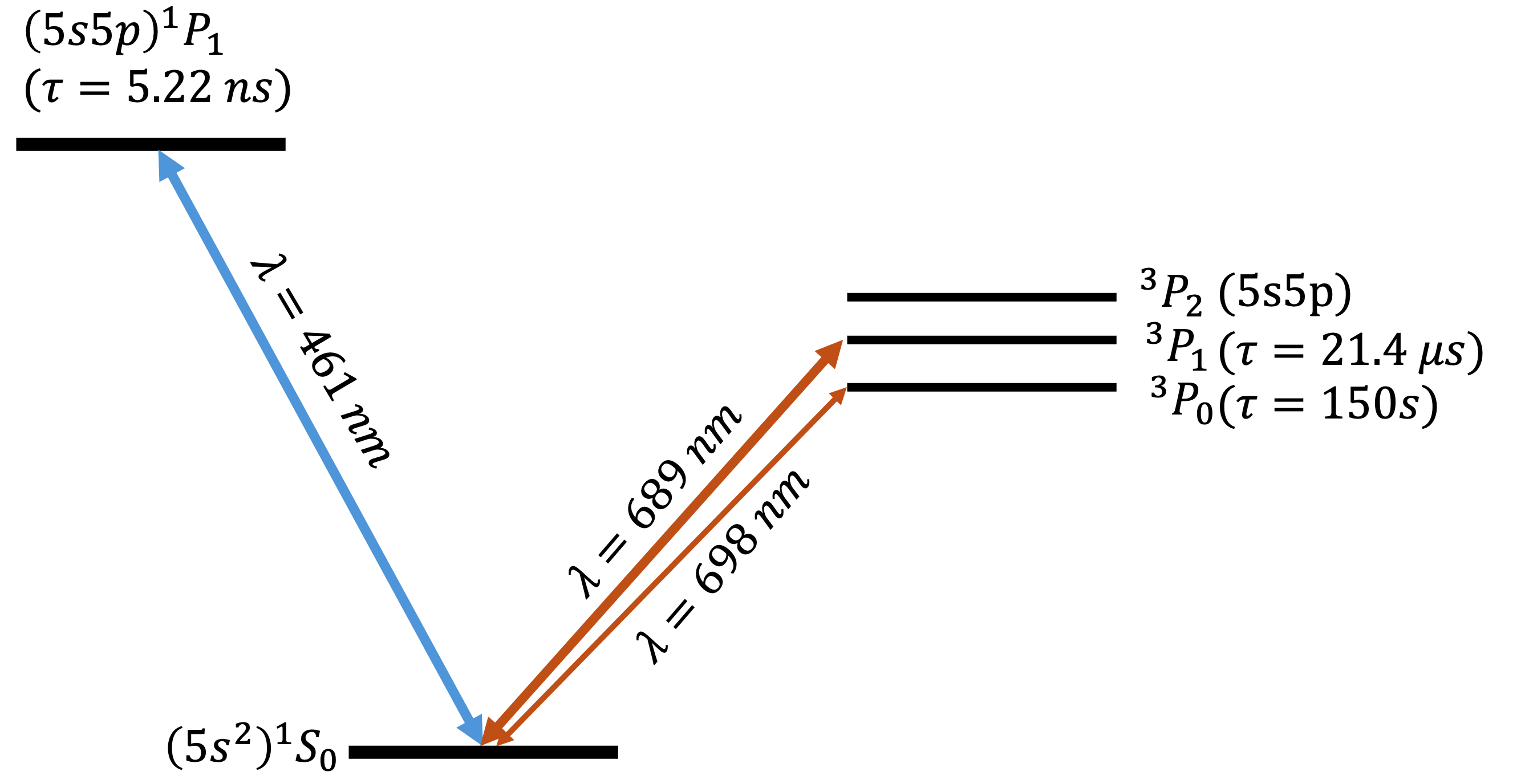}
    \caption{Schematic diagram illustrating the energy levels of the strontium atom relevant to our discussion \cite{boyd2007high}.}
    \label{fig:strontium_levels}
\end{figure}

One promising implementation involves using the narrow optical transitions at 698 nm \cite{hu2017atom, hu2019sr, baynham2025prototype} or 689 nm \cite{rudolph2020large,wilkason2022atom,wang2024robust} in strontium atoms\footnote{One can also access the long-lived excited $^{3}P_0$ state in strontium via three-photon transitions \cite{carman2025collinear}}. Figure~\ref{fig:strontium_levels} illustrates the relevant energy levels. As an example, we consider the narrower 698 nm transition here. Atoms are initialized in the ground state $^1S_0$, then coherently transferred into a superposition of $^1S_0$ and $^3P_0$ using a $\pi/2$ pulse. A Mach-Zehnder interferometer sequence is implemented such that—under ideal, fully coherent conditions—all atoms interfere constructively into the $^3P_0$ state after the $\pi/2$ pulse at the end of the sequence. Any atoms remaining in the $^1S_0$ state thus correspond to decohered or collapsed outcomes, constituting the collapsed arm.  We note that these decohered or collapsed atoms will have a 50:50 probability of being in $^1S_0$ or $^3P_0$ after the $\pi/2$ pulse.  Thus, only half of them can be removed from the superposition trap.

\begin{figure*}[ht]
    \centering
    \includegraphics[width=\linewidth]{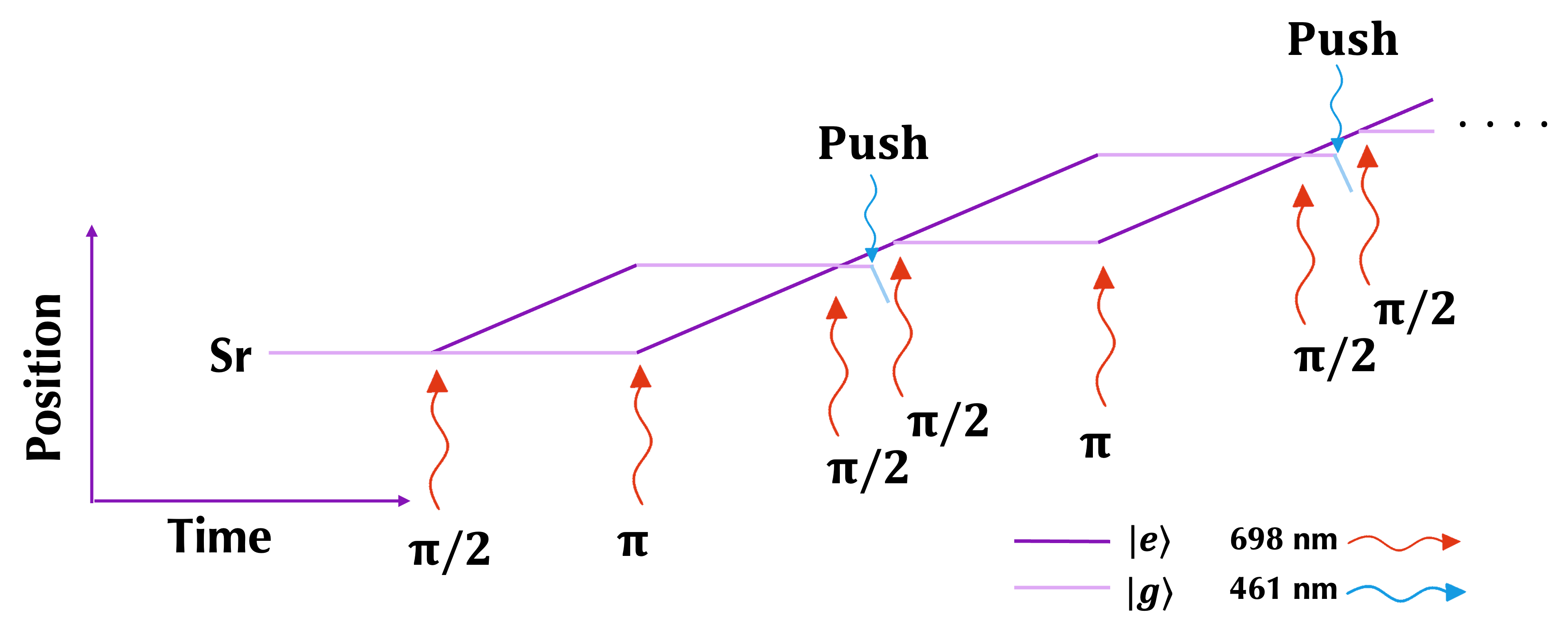}
    \caption{Illustration of the possible protocol of superposition trap implemented in atom interferometry. $\ket{g}$ represents the ground state and $\ket{e}$ represents the excited state.
    The coherent atoms are contained within the physical bounds using $698$ nm laser pulses, while half of the collapsed atoms can be pushed using the $461$ nm laser pulses. As these collapsed atoms are physically separated, the method allows us to measure them in real time without impacting the coherent atoms.}
    \label{fig:interferometer_schematic}
\end{figure*}

To isolate these atoms, the $461$ nm transition between $^1S_0$ and $^1P_1$ can be used to selectively push atoms in the $^1S_0$ population, deflecting them from the main interferometric region \cite{rudolph2020large,wang2024robust}. The atoms in $^3P_0$ remain mostly unaffected. An additional $\pi/2$ pulse then begins a new interferometer, and the entire process can be repeated many times. Figure~\ref{fig:interferometer_schematic} shows a schematic representation of this protocol. This protocol has resemblance to the strontium multi-loop interferometer demonstrated in \cite{wang2024robust}, though is not identical due to the extra $\pi/2$ and 461 nm push pulses.

Atom interferometers also provide the added advantage of pulse sequence programmability, enabling flexible control over the spacetime paths of the atomic wave packets. This allows experimental access to a broad class of trajectories and system parameters, offering a rich testing ground to explore the dependence of collapse or decoherence rates on experimental conditions.

\section{Discussion}
The superposition trap introduces a tool wherein quantum coherence becomes not just a property of the system, but a physical parameter used for dynamical control. By constructing regions of space or Hilbert space where interference conditions lead to self-confinement, we can spatially separate coherent and incoherent components of a quantum system ensemble. This strategy circumvents the conventional dilemma of measurement-induced collapse by focusing on the leakage of incoherent components in an ensemble, which need not rely on measuring the trapped part of the ensemble at all.

Such a technique opens multiple lines of inquiry. First, it provides a direct method to monitor decoherence timescales in real time in open quantum systems. Second, it creates a new tool to place experimental bounds on spontaneous collapse models such as CSL. For example, the CSL model predicts a collapse rate that depends on mass and spatial separation \cite{bassi2003dynamical,bassi2013models,ghirardi1990markov}. A superposition trap implemented in an atom interferometer with well-controlled parameters—such as separation between arms, atom number, and interaction time—could provide data to constrain this rate.

The trap concept also raises questions about the limits of interference-based confinement. In practical setups, destructive interference is never perfect due to imperfections in alignment, pulse fidelity, and environmental noise. Thus, a careful theoretical and experimental study of leakage dynamics in realistic superposition traps is needed. This includes investigating the role of phase noise, thermal decoherence effects, and interactions in many-body systems, all of which can mimic or mask true decoherence events.

Another avenue involves adapting the superposition trap for use in other systems such as photonic systems, superconducting qubits, or hybrid quantum systems. While our proposed atom-interferometric implementation is conceptually natural and experimentally mature, the underlying principle is general and may inspire analogous architectures in other quantum platforms.

\section{Conclusion}
% Summarize your contribution, future directions.
We have introduced the concept of a superposition trap—an interferometric or spatial configuration that selectively confines coherent quantum states while permitting decohered or collapsed states to escape. This idea leverages the continuity equation in quantum mechanics as a filtering mechanism, turning coherence into a physically observable degree of freedom. Implemented in an atom interferometer, this mechanism allows for real-time monitoring of collapse or decoherence events without requiring direct measurement on the quantum system in the trap during its coherent evolution.

Such a setup has the potential to provide new experimental evidence regarding the nature and timescale of wavefunction collapse. By selectively detecting leakage from the superposition trap, we gain access to collapse dynamics that are otherwise hard to access in traditional measurement-based protocols. This not only has implications for foundational questions in quantum mechanics but might open possibilities for new quantum control schemes and error-filtering architectures in quantum information science.

Future work will focus on the detailed modeling of decoherence dynamics within the trap, quantifying leakage rates under various environmental couplings, and optimizing the interferometric parameters to enhance sensitivity to potential collapse-inducing mechanisms. We envision that superposition traps may eventually serve as both diagnostic and control tools in precision quantum experiments, offering an elegant and physically motivated probe of the quantum-classical boundary.

\begin{acknowledgments}
% Acknowledge any funding, collaborators, or helpful discussions.
Hardeep Singh extends heartfelt thanks to Shafaq Elahi, in conversation with whom the seed of this work was first conceived. He also would like to thank André de Gouvêa for engaging in conversations that helped refine some ideas. We acknowledge support from NSF award PHY-2409710.
\end{acknowledgments}

\appendix
\section{Generality of Continuity in Quantum using Path Integral} \label{Appendix: continuity path}

The section \ref{sec: continuity} discussed how the continuity equation is embedded within the Schrödinger equation. However, one may ask whether continuity is a special feature of the Hamiltonians typically considered in simple systems, or if it is a general property of quantum theory itself. Since the continuity equation is not immediately apparent in the wavefunction formalism—requiring careful manipulations of the probability density and current—it is natural to seek a more general perspective. In this appendix, we show that the Feynman path integral formalism provides such an argument, establishing the general emergence of the continuity equation in position space.

As a brief recap, in the path integral framework each possible trajectory contributes to the wavefunction amplitude with a weight factor
\begin{equation}
    e^{\frac{i}{\hbar} S[x(t)]}
\end{equation}
where $S[x(t)]$ is the action functional associated with the path $x(t)$. (We set $\hbar=1$ in what follows.) To uncover the essence of continuity in this setting, we must demonstrate that paths crossing a point in space–time with vanishing amplitude necessarily cancel out. This implies that contributions from paths intersecting a null boundary are zero, thereby trapping probability density inside such a boundary.
\begin{figure}
    \centering
    \includegraphics[width=\linewidth]{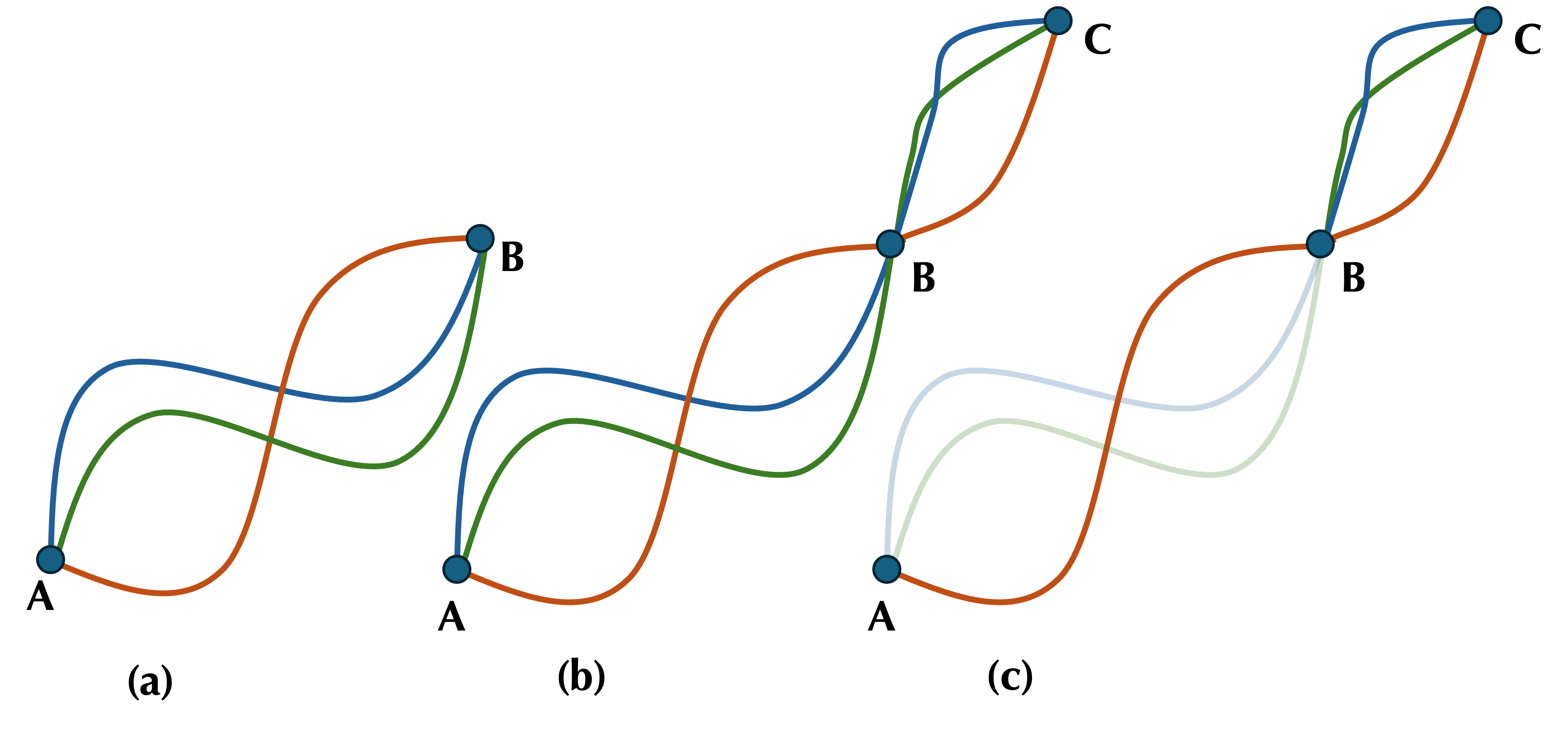}
    \caption{(a) illustrates three of the possible infinite paths that the particle can take from point $A$ to be observed at point $B$.\\ (b) illustrates the paths that go from $A$ to $C$ and pass through $B$, contributing to the function $\psi_{A\rightarrow B\rightarrow C}$. (c) Schematic illustration of the factorization from Eq.~\ref{eqn: ABC single sum} to Eq.~\ref{eqn: ABC double sum}.  For every path segment $A\!\to\!B$ there exists an infinite set of independent continuations $B\!\to\!C$.  Since the $B\!\to\!C$ paths are available for each $A\!\to\!B$ trajectory, the amplitude for $A\!\to\!C$ can be reorganized as a product of two independent sums: one over all $A\!\to\!B$ paths and one over all $B\!\to\!C$ paths.}
    \label{fig:Path A to B}
\end{figure}
Consider a particle beginning at position $A$ and observed at position $B$ [see Fig. \ref{fig:Path A to B} (a) which illustrates three of the possible infinite paths]. In continuous space, infinitely many paths connect $A$ to $B$, each with a corresponding action $S[x_i(t)]$. The (non-normalized) amplitude at $B$ is therefore
\begin{equation}
\psi_{A\rightarrow B} = \sum^\infty_i e^{i S[x_i(t)]_{AB}}
\end{equation}
Now introduce a third point $C$ in space–time. The amplitude for the particle to propagate from $A$ to $C$ is
\begin{equation}
\psi_{A\rightarrow C} = \sum^\infty_i e^{i S[x_i(t)]_{AC}}
\end{equation}
We then decompose this amplitude according to whether the intermediate paths pass through $B$:
\begin{equation}
\psi_{A \rightarrow C} = \psi_{A \rightarrow B \rightarrow C}\, +\, \psi_{A \rightarrow \cancel{B} \rightarrow C}
\end{equation}
That is, the amplitude to reach $C$ is the sum of amplitudes for trajectories passing through $B$ and for those that do not.

Focusing on the first contribution,
\begin{equation}
\begin{split}
\psi_{A \rightarrow B \rightarrow C } &= \sum^\infty_j e^{i S[x_j(t)]_{ABC}}\\
&= \sum^\infty_j e^{i \big( S[x_j(t)]_{AB} + S[x_j(t)]_{BC}\big)}
\end{split}
\label{eqn: ABC single sum}
\end{equation}
Because the number of paths is infinite, we may reorganize the sum: for each path from $A$ to $B$, there exist infinitely many continuations from $B$ to $C$. Figure \ref{fig:Path A to B} (b) and (c) illustrates the argument pictorially. Thus,
\begin{equation}
\psi_{A \rightarrow B \rightarrow C } = \sum^\infty_i \left[ e^{i S[x_i(t)]_{AB}} \, \sum^\infty_j \left[ e^{i S[x_j(t)]_{BC}} \right] \right]
\label{eqn: ABC double sum}
\end{equation}
This factorizes into
\begin{equation}
\begin{split}
\psi_{A \rightarrow B \rightarrow C } & = \sum^\infty_i \left[ e^{i S[x_i(t)]_{AB}} \right]  \sum^\infty_j \left[ e^{i S[x_j(t)]_{BC}} \right]\\
& = \psi_{A \rightarrow B} \, \psi_{B \rightarrow C}
\end{split}
\end{equation}

Hence, the contribution of the paths passing through $B$ depends multiplicatively on the amplitude at $B$. If the amplitude at $B$ vanishes, then $\psi_{A \rightarrow B \rightarrow C} = 0$ regardless of the details of the continuation from $B$ to $C$. This leads to the notion of a probability bubble: if a closed surface in space–time consists entirely of points with vanishing amplitude, then paths on one side of the surface make no contribution to amplitudes on the other side.

Crucially, no assumptions were made about the particular form of the action. The only mathematical condition invoked was the additivity of the action,
\begin{equation}
S[x_j(t)]_{ABC} = S[x_j(t)]_{AB} + S[x_j(t)]_{BC}
\end{equation}
which holds generally when the action is defined as
\begin{equation}
S[x(t)] = \int_{t_1}^{t_2} L[x(t), \dot{x}(t), t]\, dt,
\end{equation}
since the integral is additive over adjoining intervals. Thus, the argument establishes the generality of the continuity constraint in quantum theory, independent of any specific Hamiltonian, Lagrangian, or Action.

\bibliographystyle{apsrev4-2}
\bibliography{collapse_trap_refs}

\end{document}